\documentclass[
superscriptaddress,
 amsmath,amssymb,
 aps,
]{revtex4-2}

\usepackage{units}
\usepackage{graphicx}
\expandafter\let\csname equation*\endcsname\relax
\expandafter\let\csname endequation*\endcsname\relax

\newcommand{\comment}[1]{}

\begin{document}

\title{The emergence of macroscopic currents in photoconductive sampling of optical fields}

\author{Johannes Sch{\"o}tz}
\email{johannes.schoetz@mpq.mpg.de}
\affiliation{Department of Physics, Ludwig-Maximilians-Universit\"at Munich, D-85748 Garching, Germany}
\affiliation{Max Planck Institute of Quantum Optics, D-85748 Garching, Germany}
\author{Ancyline Maliakkal}
\affiliation{Department of Physics, Ludwig-Maximilians-Universit\"at Munich, D-85748 Garching, Germany}
\affiliation{Max Planck Institute of Quantum Optics, D-85748 Garching, Germany}
\author{Johannes Bl{\"o}chl}
\affiliation{Department of Physics, Ludwig-Maximilians-Universit\"at Munich, D-85748 Garching, Germany}
\affiliation{Max Planck Institute of Quantum Optics, D-85748 Garching, Germany}
\author{Dmitry Zimin}
\affiliation{Max Planck Institute of Quantum Optics, D-85748 Garching, Germany}
\author{Zilong Wang}
\affiliation{Department of Physics, Ludwig-Maximilians-Universit\"at Munich, D-85748 Garching, Germany}
\affiliation{Max Planck Institute of Quantum Optics, D-85748 Garching, Germany}
\author{Philipp Rosenberger}
\affiliation{Department of Physics, Ludwig-Maximilians-Universit\"at Munich, D-85748 Garching, Germany}
\affiliation{Max Planck Institute of Quantum Optics, D-85748 Garching, Germany}
\author{Meshaal Alharbi}
\affiliation{Attosecond Science Laboratory, Physics and Astronomy Department, King Saud University, Riyadh 11451, Saudi Arabia}
\author{Abdallah M. Azzeer}
\affiliation{Attosecond Science Laboratory, Physics and Astronomy Department, King Saud University, Riyadh 11451, Saudi Arabia}
\author{Matthew Weidman}
\affiliation{Department of Physics, Ludwig-Maximilians-Universit\"at Munich, D-85748 Garching, Germany}
\affiliation{Max Planck Institute of Quantum Optics, D-85748 Garching, Germany}
\author{Vladislav S. Yakovlev}
\affiliation{Department of Physics, Ludwig-Maximilians-Universit\"at Munich, D-85748 Garching, Germany}
\affiliation{Max Planck Institute of Quantum Optics, D-85748 Garching, Germany}
\author{Boris Bergues}
\affiliation{Department of Physics, Ludwig-Maximilians-Universit\"at Munich, D-85748 Garching, Germany}
\affiliation{Max Planck Institute of Quantum Optics, D-85748 Garching, Germany}
\author{Matthias F. Kling}
\email{matthias.kling@lmu.de}
\affiliation{Department of Physics, Ludwig-Maximilians-Universit\"at Munich, D-85748 Garching, Germany}
\affiliation{Max Planck Institute of Quantum Optics, D-85748 Garching, Germany}

\begin{abstract}
Photoconductive field sampling is a key methodology for advancing our understanding of light-matter interaction and ultrafast optoelectronic applications. For visible light the bandwidth of photoconductive sampling of fields and field-induced dynamics can be extended to the petahertz domain. Despite the growing importance of ultrafast photoconductive measurements, a rigorous model for connecting the microscopic electron dynamics to the macroscopic external signal is lacking. This has caused conflicting interpretations about the origin of macroscopic currents. Here, we present systematic experimental studies on the macroscopic signal formation of ultrafast currents in gases. We developed a theoretical model based on the Ramo-Shockley-theorem that overcomes the previously introduced artificial separation into dipole and current contributions. Extensive numerical particle-in-cell (PIC)-type simulations based on this model permit a quantitative comparison with experimental results and help to identify the roles of electron scattering and Coulomb interactions. The results imply that most of the heuristic models utilized so far will need to be amended. Our approach can aid in the design of more sensitive and more efficient photoconductive devices. We demonstrate for the case of gases that over an order of magnitude increase in signal is achievable, paving the way towards petahertz field measurements with the highest sensitivity.
\end{abstract}

\maketitle

\section*{Introduction}
Intense few-cycle pulses can induce conductivity within a fraction of an optical cycle, enabling the ultrafast manipulation of electric currents. Such currents have been measured in various materials\cite{Schiffrin2013FieldInducedCurrents, Paasch-Colberg2016GaN, Kim2016CaF2, Kim2016Semimetalization, Mikkelsen2020OpticaGaN} and 2D-structures\cite{Heide2018PRL}. Using sub-cycle current injection as a temporal gate, in a generalization of the concept of THz photoconductive field-sampling\cite{auston1984picosecond, Castro-Camus2016PhotoconductiveSwitch}, the field-resolved measurement of optical waveforms up to PHz-frequencies has been demonstrated\cite{Sederberg2020NatComm}. For the investigation of field-dependent processes, knowledge of the carrier-envelope phase (CEP) is needed for an unambiguous determination of the field. Conventional characterization techniques, including frequency-resolved optical gating (FROG)\cite{Kane1993}, spectral-phase interferometry for direct electric-field reconstruction (SPIDER)\cite{Walmsley1999}, and the dispersion-scan (D-scan) technique \cite{Miranda12}, do not give access to the CEP. For optical frequencies, techniques such as attosecond streaking \cite{Goulielmakis_2004_first_streaking} and the stereo-ATI phase meter \cite{Rathje12} can be used to determine the electric field waveform and carrier-envelope phase, respectively. They require, however, complex ultra-high vacuum setups, in which electron time-of-flight spectra can be recorded. This has been limiting their widespread application in many laboratories.

With photoconductive field sampling in solids, a much simpler solution has been presented \cite{Sederberg2020NatComm}. Recently, the same concepts have been applied in air for the measurement of the CEP and electric field\cite{Kubullek2020CEPtag, Korobenko2020fsStreaking, Weidman2021NPSair}, which offers additional advantages: they are easier to use since they eliminate the requirement of sample fabrication and inherently use a refreshable target. Moreover, the microscopic response at the atomic/molecular level can be numerically calculated using the time-dependent Schr\"odinger equation (TDSE), and may be modeled and interpreted in the framework of the strong-field approximation (SFA) and the simpleman's model (SMM)\cite{Corkum93threestep}. The general experimental setup of strong-field sub-cycle controlled currents is almost identical to broadband THz generation in gases\cite{Cook2000THzfield, Roskos2004THzemission, Roskos2006THzCEP} and both processes are expected to be closely linked.

Despite the importance of a detailed understanding of the signal formation in ultrafast current measurements, the macroscopic aspects were either not discussed or not fully understood to date. There is no rigorous model that connects the microscopic single electron dynamics to the macroscopic current signal that is measurable on the electrodes. The macroscopic aspects of signal formation were largely ignored even though processes such as electron scattering are expected to play an important role\cite{Korobenko2020fsStreaking,Weidman2021NPSair}. Apart from Refs.\,\cite{Boolakee2020grapheneDistance, keiber2016novel} where the role of the electrode distance was investigated, no systematic studies exist. 

\begin{figure}[htbp!]
	\centering\includegraphics[width=4in]{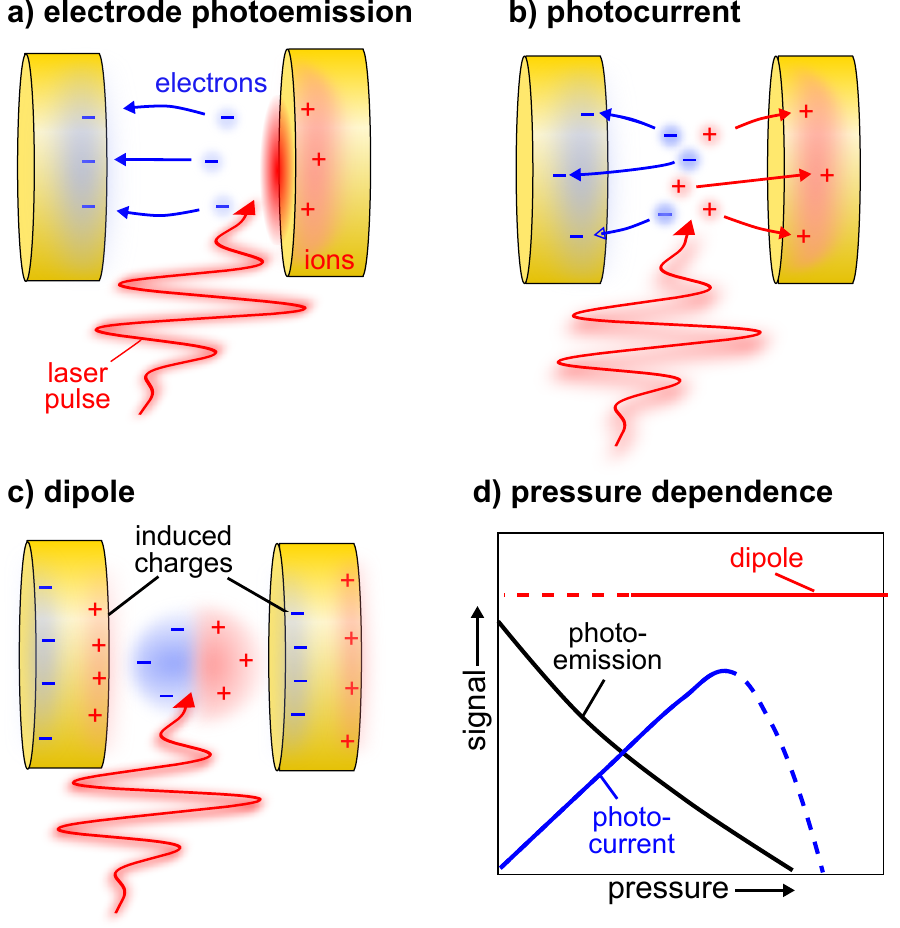}
	\caption[Heuristic models]{\label{Fig_heuristic_models} Heuristic models used in the interpretation of macroscopic signal formation in ultrafast current sampling in gases: a) electrode photoemission, b) photocurrent, and c) dipole. The electrodes are depicted in yellow, where individually a current towards ground can be measured. d) Expected pressure dependence for the mechanisms depicted in a)-c), where Coulomb interaction is neglected.}
\end{figure}

An overview over existing, purely heuristic, models for macroscopic signal generation in ultrafast current sampling in gases, together with the expected dependence on pressure is shown in Fig.\,\ref{Fig_heuristic_models}. Note that all these rather simple models ignore the Coulomb interaction between light-created charges. In the first model, a few-cycle optical laser pulse can directly induce CEP-dependent strong-field photoemission from one of the electrodes, resulting in a corresponding current (cf. Fig.\,\ref{Fig_heuristic_models}a)). Here, the sign of the current would be opposite if the counter-electrode is illuminated. Generally, this contribution is expected to decrease from vacuum towards higher pressures (black solid line in Fig.\,\ref{Fig_heuristic_models}d)) due to scattering-limited propagation of the released electrons. In most ultrafast current sampling experiments electrode photoemission produces a background that is avoided. The second heuristic model is denoted as photocurrent and shown in Fig.\,\ref{Fig_heuristic_models}b). Here, not the electrodes but the gas medium between them is (tunnel)ionized. The same laser field then accelerates ions and free electrons towards the electrodes. The current depends on the laser field, e.g. for a certain CEP more electrons impinge on the left detector than the right (as shown in Fig.\,\ref{Fig_heuristic_models}b)) or vice versa when introducing a shift of $\pi$ in CEP. The blue solid line in Fig.\,\ref{Fig_heuristic_models}d) shows the expected pressure dependence. At low pressures, where the mean-free path is larger than the distance to the electrodes, the number of detected charges is proportional to the medium density and the current grows linearly with pressure. For high enough pressures, charges cannot reach the detector anymore before scattering and in turn losing their strict relationship to the CEP. As a result, the macroscopic current would be expected to drop sharply. Previous work discussed whether this contribution can play a role at ambient pressure \cite{Korobenko2020fsStreaking}. The third heuristic model involves a laser-induced charge dipole following the ionization of the medium, cf. Fig.\,\ref{Fig_heuristic_models}c). This induces an image charge on the electrodes, yielding a macroscopic current. Interestingly, in this case, a rather constant current over pressure is expected as depicted as solid red line in Fig.\,\ref{Fig_heuristic_models}d), since the total charge scales linearly with pressure, while the mean-free path scales inversely, keeping the dipole strength constant. While these heuristic models have been invoked in the interpretation of previous results \cite{Kim2018TipToe,Korobenko2020fsStreaking,Weidman2021NPSair}, their validity has been debated and they failed to provide quantitative predictions. Clearly, to further advance our understanding of ultrafast photoconductive sampling in gases, a quantitative model for the macroscopic signal formation is crucial.

Here, we present a newly developed rigorous theoretical approach based on the Ramo-Shockley theorem (cf. Refs.\,\cite{He2001ShockleyRamo, Riegler2019RSTheoremExtension, Ramo1939RSTheorem, Shockley1938RSTheorem}), which resolves controversies originating from the previously used artificial separation into photocurrent and dipole contributions. Our extensive numerical particle-in-cell (PIC)-type simulations based on this model enable a quantitative comparison with the experimental results and identify the roles of electron scattering and Coulomb interactions. The model overcomes the limitations of the existing heuristic models, and it provides a new fundamental understanding of photoconductive sampling of optical fields. The ability of the model for quantitative predictions paves the way towards the implementation of more sensitive and more efficient petahertz optoelectronic devices.

\begin{figure}[htbp!]
	\centering\includegraphics[width=6.5in]{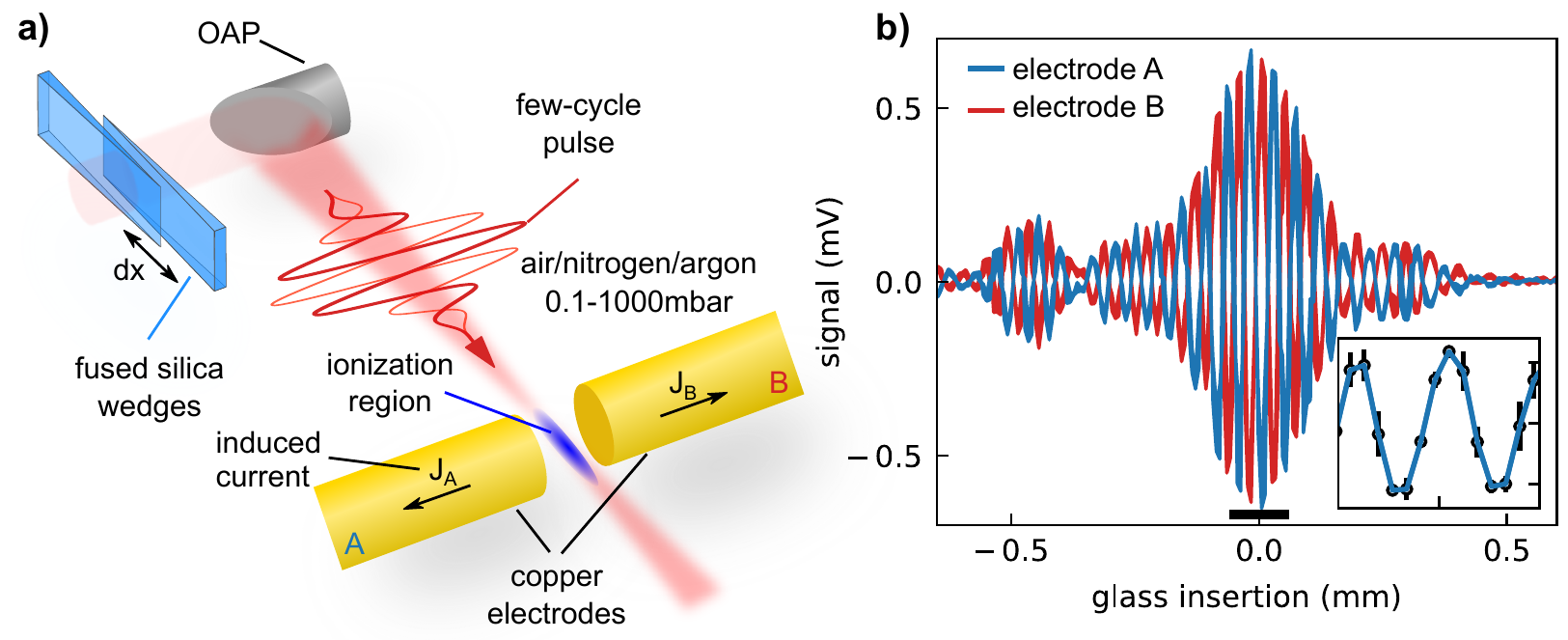}
	\caption[Experimental setup and theoretical approach]{\label{Fig_gas_experimental_setup} Laser-field induced currents in gases: a) Experimental setup (see text for details). OAP: off-axis parabola. b) Current dispersion scan trace obtained by moving the fused silica wedges when recording currents from electrode A (blue line) and B (red line). The line thickness corresponds to the standard deviation of three consecutive measurements. The inset shows a zoomed view (black bar in b) of the signal from electrode A with individual datapoints marked as dots.}
\end{figure}

\section*{Results}
In our experiment (see Methods for details), laser pulses are focused into a gas where they reach intensities of $10^{14}\,\mathrm{W/cm}^2$ and create a partially ionized ensemble of atoms and molecules. The setup is shown in Fig.\,\ref{Fig_gas_experimental_setup}a). Fused silica wedges are used to control the dispersion. The laser pulses of 4.5\,fs duration at 750\,nm are focused by an off-axis parabola (OAP, f=101.6\,mm) to a spot size below 10\,$\mu$m, as determined using a CCD camera. Additionally, a mirror-based telescope can be introduced in front of the setup to increase the spot size by about a factor of three. The focused laser creates an ionization region between two electrodes. The propagation of ionized electrons induces currents on the electrodes, which are produced from a copper wire with a diameter of approximately 500\,$\mu$m. The electrodes, denoted A and B in Fig.\,\ref{Fig_gas_experimental_setup}a), are mounted individually on computerized stages, permitting distance control. The electrode assembly is placed on a linear closed-loop 3D-stage for fine positioning with respect to the ionization region. The positioning is monitored by an in-situ imaging system installed downstream. The currents measured between each of the electrodes and ground are amplified by two transimpedance amplifiers (Femto DLPCA-200) with a gain of $10^9\mathrm{V/A}$. The resulting voltage pulses are detected via a two-channel lock-in amplifier (Z\"urich Instruments HF2LI). The focusing mirror and electrodes are placed in a vacuum chamber which is used to change the gas species (air, nitrogen and argon) and vary the pressure (0.1 to 1000\,mbar) in the ionization region. An example signal trace, obtained by scanning the dispersion with the wedges is depicted in Fig.\,\ref{Fig_gas_experimental_setup}b) for both electrodes. The observed oscillations are caused by the change of the CEP with the dispersion, while the envelope reflects the change in pulse duration and peak intensity. Since the electrodes A and B measure the current in opposite directions, their signals are $\pi$ out of phase.

Our theoretical model is explained in detail in the Methods section. Briefly, the calculations are based on the Ramo-Shockley theorem \cite{Ramo1939RSTheorem, Shockley1938RSTheorem}. Here, the induced charge $Q$ and current $I$ on the electrode, caused by a particle with charge $q$ at position $\vec{r}$ and velocity $\vec{v}$, are given by\cite{He2001ShockleyRamo, Riegler2019RSTheoremExtension}:
\begin{align}
	Q&=-q \phi_\mathrm{0}(\vec{r}),\\
	I&=q\vec{v}\vec{E_\mathrm{0}}(\vec{r}),
\end{align}
where $\phi_0$ is the weighting potential and $\vec{E}_0$ is the weighting field. For any arrangement of electrodes, the weighting potential can be calculated by setting the potential on the electrode under consideration to unity (1\,V in SI units) and to zero on all other electrodes. For an ensemble of charges, the induced charge is given by the sum over the individual particle contributions.

\begin{figure}[htbp!]
	\centering\includegraphics[width=6.5in]{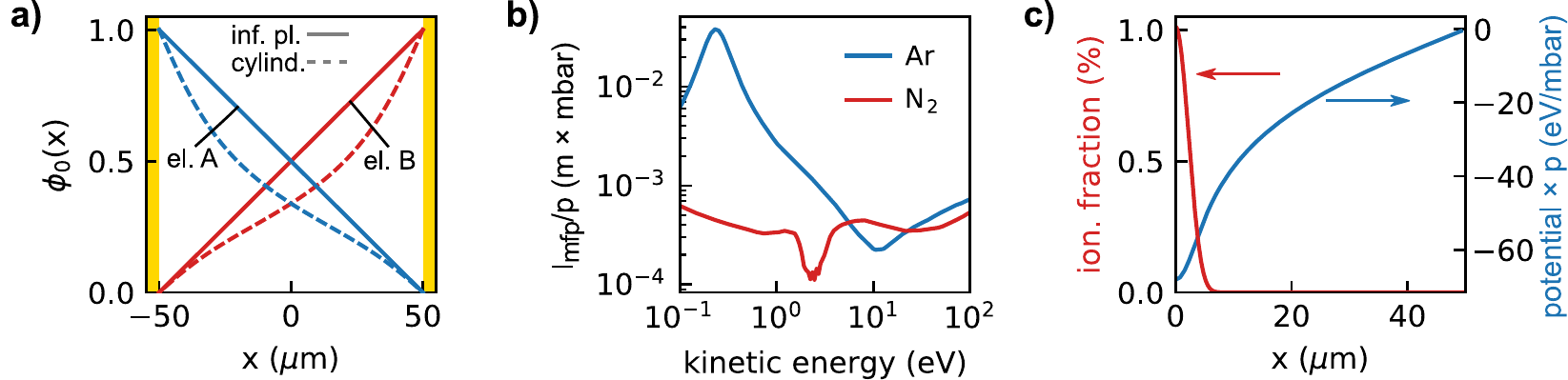}
	\caption[Experimental setup and theoretical approach]{\label{Fig_gas_theory} Theoretical modeling: a) Ramo-Shockley weighting potentials for an electrode distance of 100\,$\mu\mathrm{m}$. b) Electron mean-free path at 1\,mbar for argon and nitrogen as a function of kinetic energy. c) Ionization fraction and electrostatic potential of the ion background for 1\,mbar argon. Simulation parameters: $\omega_\mathrm{0}=25\,\mu\mathrm{m}$ and $I=1.6\cdot10^{14}\mathrm{W/cm}^2$.}
\end{figure}

For infinitely extended parallel plates, the weighting potential can be obtained analytically and has a very simple form: It depends on the position of the charge between the electrodes. It is one at the electrode under consideration and linearly decays to zero at the other electrode. We use this idealized weighting potential in our simulations due to its simplicity. For comparison, we numerically calculated the weighting potential for a realistic geometry of two opposing metallic cylinders with a ratio of their distance to the radius of 8. The electrodes were meshed using GMSH\cite{geuzaine2009gmsh, gmshprogram} and the electrostatic calculation was performed using the boundary-element implementation of scuff-em\cite{SCUFF1,SCUFF2}. The calculated weighting potential along the cylinder axis for electrode A (blue dashed line) is shown in Fig.\,\ref{Fig_gas_theory}a). Compared to the linear infinite plate solution (blue solid line), the realistic weighting potential decays faster when moving away from the electrode. The weighting potentials for electrode B (red dashed and solid lines) are symmetric around the center plane at $x=0$.

For the numerical simulations of our experiments, an electrostatic particle-in-cell (PIC) code was developed. After emission by the laser, the electron propagation under the influence of scattering and electric fields is considered. The laser is modeled with a Gaussian envelope in space and time. The duration of the pulse was 4.5\,fs full-width-at-half-maximum (FWHM) of the intensity envelope. In our Monte-Carlo approach, the initial positions of charges is randomly sampled from the radially-resolved ionization fraction and the emission time from the tunneling rate. The propagation of charges is performed via the Velocity-Verlet algorithm\cite{Swope1982VelocityVerlet}. For each time step in the propagation, the electron-neutral (atom or molecule) scattering probability is calculated via the mean-free path $l_\mathrm{mfp}$ and Monte-Carlo sampling. The mean-free paths for argon (blue line) and nitrogen (red line) used in the simulations are shown in Fig.\,\ref{Fig_gas_theory}b) at 1\,mbar. Figure\,\ref{Fig_gas_theory}c) shows the ionization fraction (red line) and the electrostatic potential of the ion background for 1 mbar argon (blue line) from the center of the simulation towards one electrode. The strength of the ion potential also explains why related experiments that measured the generated charge via bias voltages, were conducted at low pressures\cite{Kim2018TipToe} or had to apply kV-level biases.

\begin{figure}[htbp!]
	\centering\includegraphics[width=6.in]{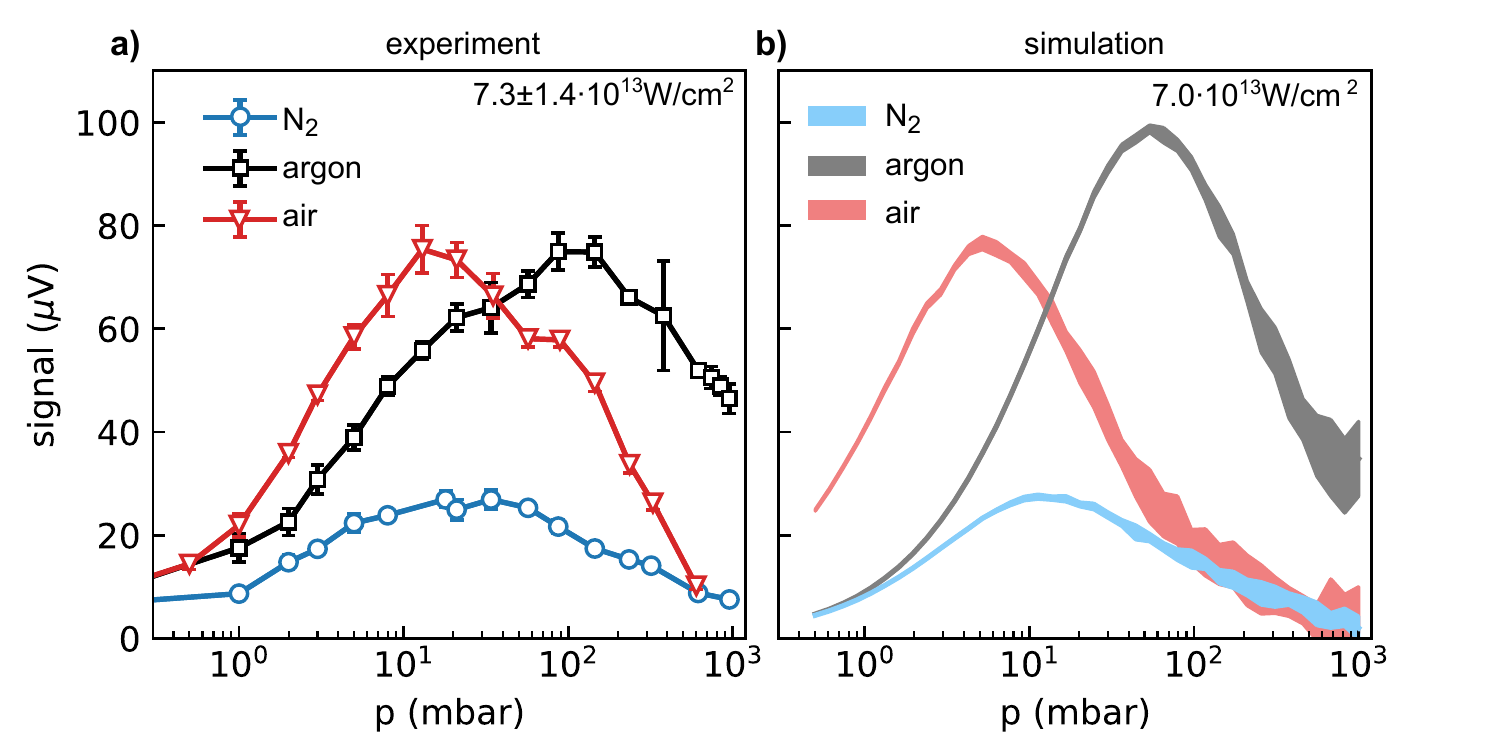}
	\caption[Experimental pressure-dependence]{\label{Fig_gas_pressure_results} Pressure-dependence of the maximum current signal amplitude: a) Experimental data for different gases. b) Simulation results ($\omega_0$=8\,$\mu$m).}
\end{figure}

The recorded pressure dependences of the maximum current signal amplitudes for nitrogen (blue line), argon (black line), and air (red line) are shown in Fig.\,\ref{Fig_gas_pressure_results}a). The experimental data has been averaged over three individual dispersion scans (see Fig.\,\ref{Fig_gas_experimental_setup}b), and the error bars correspond to the standard deviation. Performing the measurements via dispersion scans is necessary, since increasing the pressure leads to a shift of the maximum to lower glass insertions. The electrode distance was around 100\,$\mu$m. A rather low intensity of $7.3\cdot 10^{13}\mathrm{W/cm}^2$ was chosen in order to avoid reshaping of the dispersion trace with increasing pressure. Starting from low pressure, all three curves are increasing and reach a maximum at different pressures, nitrogen at around 30\,mbar, argon at 100\,mbar, and air at 10\,mbar. They subsequently decay going towards 1000\,mbar. The maximum signal amplitude in argon and air is roughly equal, whereas it is around a factor of three lower for nitrogen.

The simulations reproduce the main features of the pressure dependence well (cf. Fig.\,\ref{Fig_gas_pressure_results}b), especially the relative maximal amplitudes and pressures at which the maximum is reached. An intensity of $7\cdot10^{13}\mathrm{W/cm}^2$, close to the experimental conditions, provides the best results from the simulations. Since the individual pseudo-electrons have a larger weight for higher pressure, a higher standard deviation of the Monte-Carlo simulations is obtained for higher pressures. The simulated distributions are slightly narrower than the experimental ones, which is likely due to the 2D-approximation.

\begin{figure}[htbp!]
	\centering\includegraphics[width=4.5in]{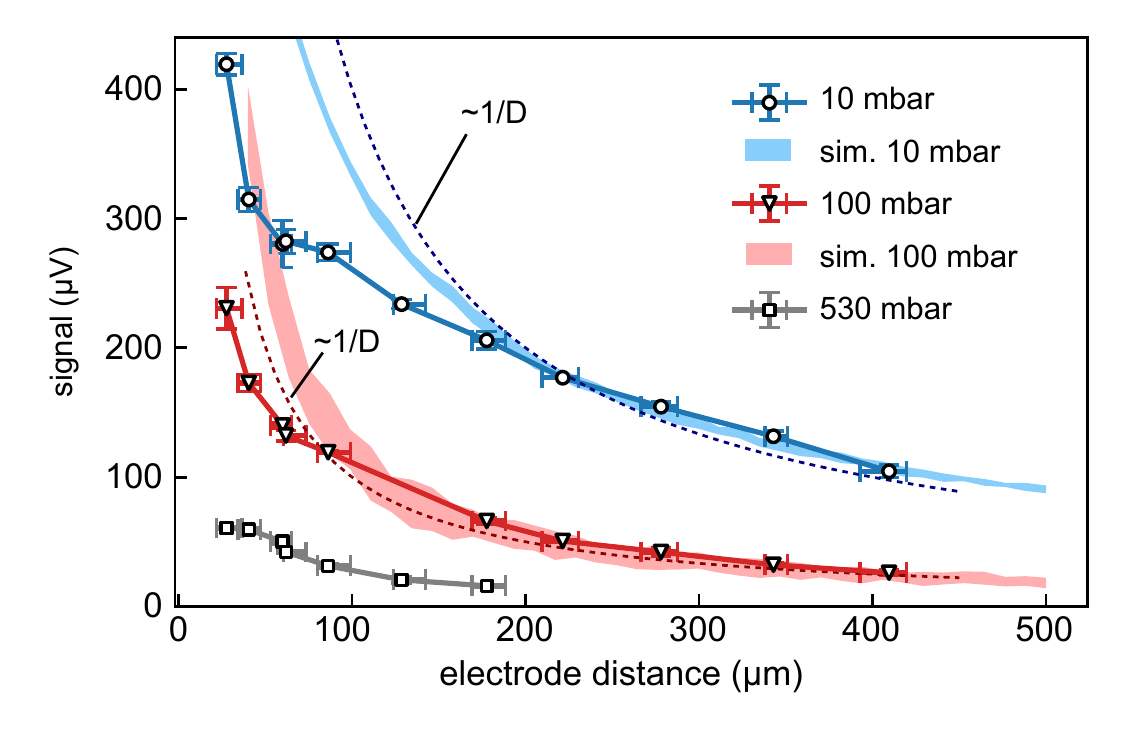}
	\caption[Electrode-distance dependence]{\label{Fig_gas_distance_experiment} Electrode-distance dependence: Maximum signal amplitudes for pressures of 10\,mbar (blue), 100\,mbar (red), and 530\,mbar (gray) in nitrogen at $8.3\pm1.8\cdot10^{13}\mathrm{W/cm}^2$. Simulation for 10\,mbar (light blue area) and 100\,mbar (light red area) for $\omega_0$\,=\,8\,$\mu$m, $I=8.0\cdot10^{13}\mathrm{W/cm}^2$. For comparison the $1/D$-dependence is shown (dashed lines).}
\end{figure}

Figure\,\ref{Fig_gas_distance_experiment} shows the electrode-distance dependence of the maximum signal amplitudes in nitrogen for pressures of 10\,mbar (blue line), 100\,mbar (red line), and 530\,mbar (gray line). As above, each data point has been obtained from the average of three dispersion scans. The signal amplitude increases nonlinearly by almost a factor of four when decreasing the distance from 420\,$\mu$m to roughly 30\,$\mu$m. At lower distances, the laser hits the electrodes, which we intentionally avoided. The simulations (light blue and light red areas, peak intensity $8\cdot10^{13}\mathrm{W/cm}^2$) reproduce the distance-dependence above roughly 150\,$\mu$m. They, however, slightly overestimate the signal for lower distances, which is further discussed below. For comparison, the $1/D$-dependence is also shown (dashed lines). It reproduces the behaviour in both experiment and simulations for larger distances.

\begin{figure}[htbp!]
	\centering\includegraphics[width=6.23in]{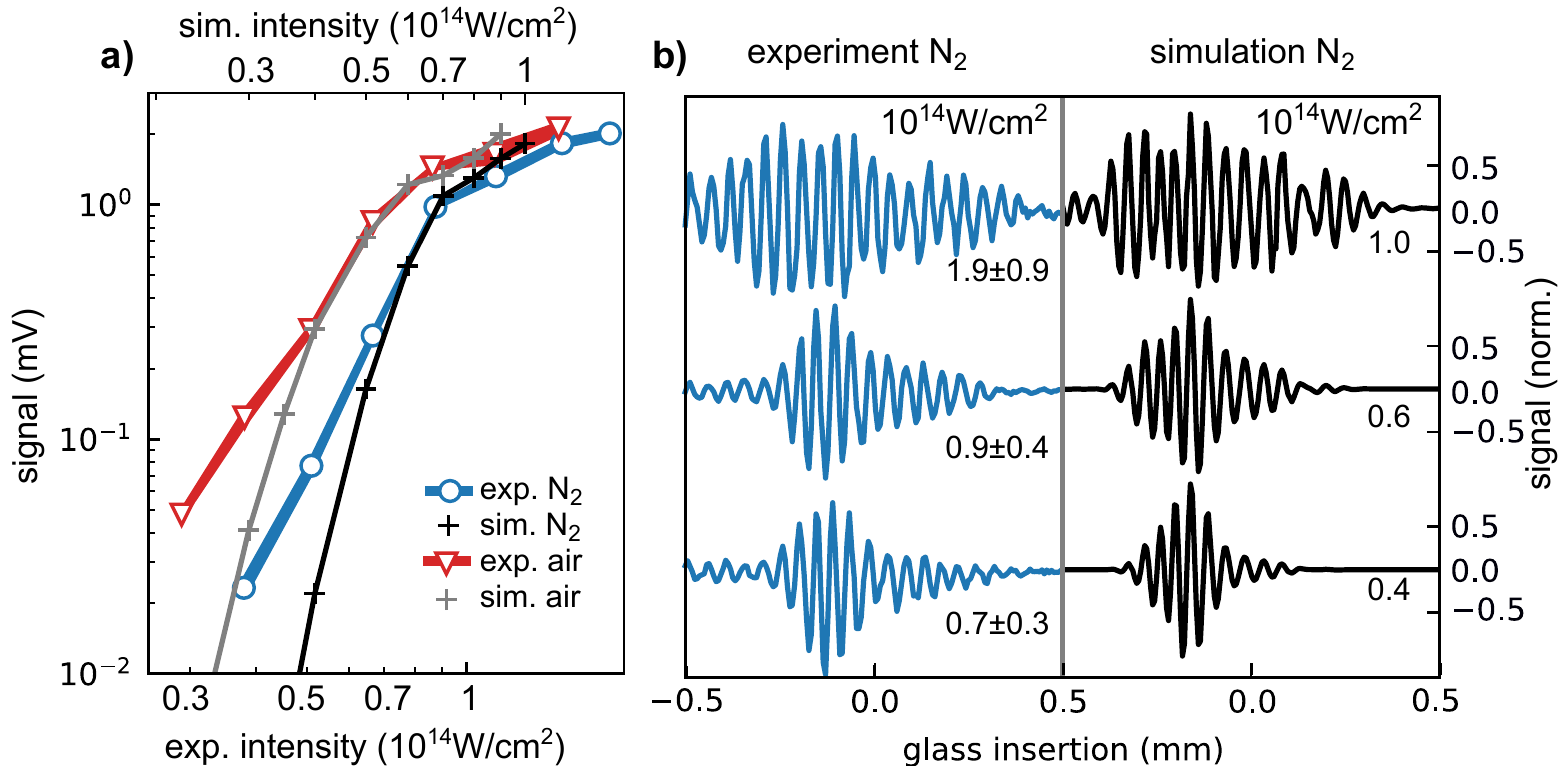}
	\caption[Intensity-dependence and signal trace reshaping]{\label{Fig_gas_intensity_experiment} Intensity-dependence and signal trace reshaping: a) Intensity-dependence of the maximum signal amplitudes for nitrogen (blue dot) and air (red triangles) together with the simulations. In the experiments, an additional telescope (see Methods) and an electrode distance of around 100\,$\mu$m was used. b) Comparison of experimental and simulated signal traces for different input intensities. The simulations used $\omega_0$\,=25\,$\mu\mathrm{m}$.}
\end{figure}

The scaling of the maximum signal strength with intensity can be seen in Fig.\,\ref{Fig_gas_intensity_experiment}a) for nitrogen (blue dots) and air (red triangles) at 25\,mbar. The signal amplitudes grow rapidly by almost two orders of magnitude when doubling the experimental peak intensity from $4$ to $8\cdot10^{13}\mathrm{W/cm}^2$. For even higher intensities, the signal amplitude saturates manifesting as a kink in the intensity vs. signal graph. In air, saturation is reached at slightly lower intensities. Below saturation, the signal amplitude in air is about a factor of 3 to 5 higher than pure nitrogen. The simulations ($\omega_\mathrm{0}=25\,\mu\mathrm{m}$) for nitrogen (black crosses) and air (gray crosses) reproduce relative amplitudes and the initial transition from the rapid increase to saturation extremely well. A small systematic offset of around 20\,\% between the experimental intensity calibration (lower axis) and the intensity in the simulation (upper axis) is observed. For the lowest intensities, the simulations underestimate the measured signal. This can be traced back to the ionization model (see Methods) not being appropriate anymore in this regime \cite{Kling2017PPT_ADKexperimental}.

To better illustrate the connection between signal saturation and the formation of the kink, the experimentally measured dispersion traces in nitrogen are shown in Fig.\,\ref{Fig_gas_intensity_experiment}b) (blue curves, left side) and compared to the simulated traces (black curves, right side). The laser pulse for the calculation is obtained from a d-scan measurement. Again, overall good agreement is observed. The low-intensity wings of the traces are underestimated in the simulation which can largely be explained by the findings above. Most importantly, the saturation of the signal trace is reproduced. It is connected to amplitude quenching at high intensities (Fig.\,\ref{Fig_gas_intensity_experiment}a). The simulations show that the occurrence of the kink is not connected to a saturation of the ionization (or the vanishing CEP-effect \cite{Kaertner2019VanishingCEP}) but is a consequence of scattering and rapidly increasing Coulomb interaction which is further discussed in the following.

\section*{Discussion}
\begin{figure}[htbp!]
	\centering\includegraphics[width=6.in]{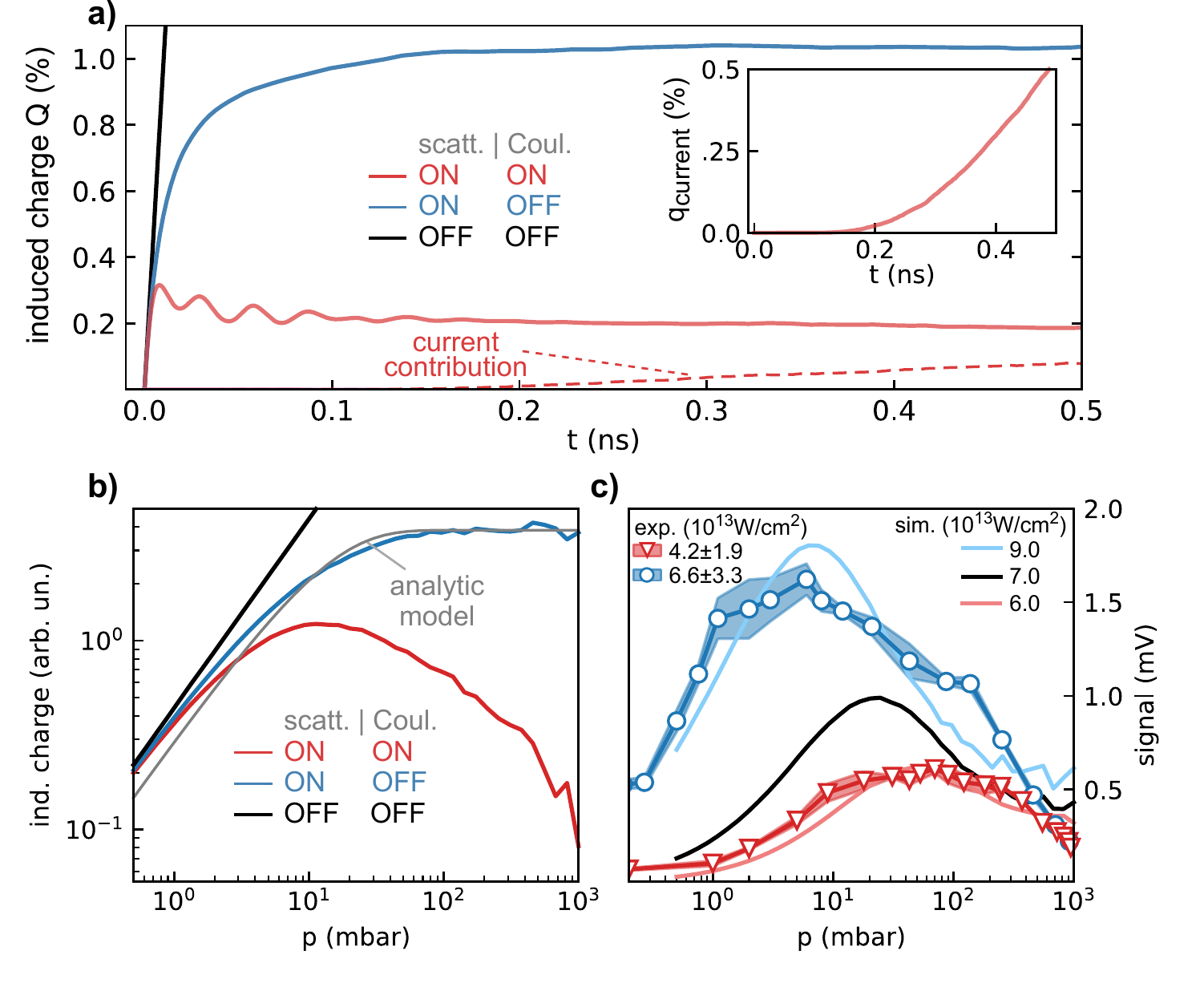}
	\caption[Role of scattering and the Coulomb-interaction]{\label{Fig_gas_scatteringCoulomb} Role of scattering and the Coulomb-interaction: a) Time-resolved induced charge $Q$ at electrode A with selectively disabled scattering and Coulomb-interaction for 100\,mbar nitrogen. The dashed line shows the current contribution to the induced charge $Q$, where the inset indicates the amount of free charges $q_\mathrm{current}$ reaching electrode A. b) Pressure dependence of the induced charge for nitrogen with selectively disabled scattering and Coulomb-interaction. c) Scaling of the pressure-dependence with different laser intensities in argon. Simulation parameters: a-b) $\omega_0$\,=8\,$\mu\mathrm{m}$ and $I=0.7\cdot10^{14}\mathrm{W/cm}^2$, c) $\omega_\mathrm{0}\,=25\,\mu\mathrm{m}$.}
\end{figure}

In order to illustrate the effect of scattering and the Coulomb interaction, we investigate their influence on the signal formation. Figure\,\ref{Fig_gas_scatteringCoulomb}a) shows the time-dependence of the induced charge on one electrode normalized by the total emitted charge for nitrogen with scattering and Coulomb interaction selectively disabled. With neither scattering nor charge interaction (black line), the relative induced charge increases rapidly and reaches around 10\,\% as would be predicted from the photocurrent. Indeed, the initial slope of all three curves is proportional to the standard expression for the photocurrent $I=\sum q\cdot v$, where the right-hand side represents the sum over all charges $q$ and their velocity $v$ in detection direction. However, when scattering is enabled (blue line), the rise of the induced charge is quickly damped and reaches close to the asymptotic value of about 1\,\% after 0.2\,ns. Qualitatively, this observation can be understood by considering that the electron propagation leads to charge induction only up to the first (isotropic) scattering event. This is because afterwards, when neglecting that the electrodes may hinder further propagation, on average no charge is induced in our model. If the charge interaction is switched on (red line), the rise is damped even faster and an asymptotic value of 0.2\,\% is reached. Here, the Coulomb interaction counteracts any charge imbalance and pulls the electrons back. Additionally, small and fast-decaying oscillations can be seen on the charge signal, that are connected to plasma oscillations. The figure also illustrates why the experimental signal is calculated from the total induced charge 0.5\,ns after the initial rise when it reaches its asymptotic value. The transient initial current burst cannot be resolved experimentally.

Additionally, we simulated the pressure-dependence with and without both effects, cf. Fig.\,\ref{Fig_gas_scatteringCoulomb}b). Without any interactions (black curve), the signal is simply proportional to the pressure. Once scattering is considered, the signal saturates above around 50\,mbar. Again, intuitively, for high pressures the contribution of a single charge is proportional to the distance it travels until the first scattering occurs, i.e. the mean-free path scaling as $1/p$, since scattering is isotropic. On the other hand, the number of charges scales linearly with $p$, therefore, the signal is constant at high pressures. Since the distance between the electron and the parent ion is limited by the electrodes, the signal drops once the mean-free path is on the order of the distance to the electrode. An analytic description of the strength of the induced charge $Q$, considering the electron distribution between the electrodes and the number of charges reaching the electrodes up to the first scattering event, is given by
\begin{equation}
Q \propto \frac{q p\cdot l_\mathrm{mfp}}{D} (1-e^{-0.5\,D/l_\mathrm{mfp}}),
\end{equation}
which is shown in Fig.\,\ref{Fig_gas_scatteringCoulomb}b) (gray line). The best fit to the simulated curve is obtained by using a mean-free path approximately twice as large as the one used in the simulation. The above expression also suggests that instead of the velocity $v_\mathrm{x}$ as for the photocurrent used by most models so far, the weighting factor for the macroscopic contribution of an individual electrons should be the effective mean free path $l_\mathrm{eff}$ in the electrode direction:
\begin{equation}
l_\mathrm{eff}=\frac{v_\mathrm{x}\cdot l_\mathrm{mfp}}{\sqrt{v_\mathrm{x}^2+v_{\perp}^2}},
\end{equation}
where $v_\mathrm{x}$ and $v_{\perp}$ is the velocity in detection direction and perpendicular to it, respectively, and where $l_\mathrm{mfp}$ is energy-dependent. These formulas should also be applicable to femtosecond streaking\cite{Korobenko2020fsStreaking}/PHz-scale nonlinear photoconductive sampling\cite{Weidman2021NPSair}.

Accounting for the Coulomb interaction makes electrons experience the attraction by the positive ion background. Intuitively, in the simplified picture above, after the first scattering event, the electron motion is not isotropic anymore but the Coulomb field acts to undo the created charge imbalance. Therefore, the contribution of an individual electron, the average electron displacement $\langle x\rangle$, falls below the mean-free path. For a given intensity, this effect becomes more pronounced at higher pressures, since the concentration of free charges and the strength of the Coulomb interaction grows proportionally to the pressure. Consequently, the measured signal is maximal at the gas pressure that maximizes $p \langle x\rangle$, which for most intensities in our study is in the range of 1 to 10\,mbar (see also Fig.\,\ref{Fig_gas_scatteringCoulomb}c).

The influence of the Coulomb interaction is further illustrated in Fig.\,\ref{Fig_gas_scatteringCoulomb}c) which shows the pressure dependence in argon measured for two peak intensities of $4.2\cdot10^{13}\mathrm{W/cm}^2$ (red triangles) and $6.6\cdot10^{13}\mathrm{W/cm}^2$ (blue dots), together with the corresponding simulations at $6\cdot10^{13}\mathrm{W/cm}^2$(light red line) and $9\cdot10^{13}\mathrm{W/cm}^2$(light blue line) and additionally at $7\cdot10^{13}\mathrm{W/cm}^2$(black line). Again, a slight systematic shift between the experimental intensity compared to the simulation is observed. In both experimental and simulation data at higher intensities, the maximum signal grows, due to the increased number of charges. The Coulomb interaction also grows, leading to a shift of the maximum to lower pressures. A similar effect is observed in Fig.\,\ref{Fig_gas_pressure_results}, when comparing the pressure dependence of nitrogen and air at equal intensity. While very similar total scattering cross-sections can be assumed for both gases, as is done in the simulation, the air contribution peaks at lower pressures since a higher number of charges is generated due to the oxygen contribution. At the same time, once the peak intensity is so high that the maximum occurs at lower pressures than where the measurement is taken, saturation occurs. This, in turn, leads to a convolution of the intensity and pressure dependence.

Regarding the signal generation mechanism, there is some debate on the roles of the asymmetric charge distribution (dipole contribution) compared to the charges that reach the electrodes (current contribution)\cite{Korobenko2020fsStreaking}. In the framework presented here, there is no real difference on the single charge level. The weighting potential (see Methods) of the electrode smoothly reaches a value of one at the electrode surface (cf. Fig.\,\ref{Fig_gas_theory}a), meaning that the total charge of the particle $q$ is induced in the electrodes regardless of whether the charge $q$ has entered the electrode or sits close to the surface. Moreover, for the idealized situation of infinite parallel plate electrodes, the induced charge $Q$ just scales linearly with decreasing distance of the particle to the electrode. In order to further clarify the roles of current and dipole contribution, we specifically looked at the amount of free charges $q_\mathrm{current}$ that reaches the electrodes, as shown in the inset of Fig.\,\ref{Fig_gas_scatteringCoulomb}a) and its' contribution to the induced charge (red dashed line, main panel). As can be seen, $q_\mathrm{current}$ constitutes a considerable fraction of the total free charge. Moreover, both the dipole contribution as well as the current contribution take part in forming the total induced charge $Q$ that is measured in the end.

Realistic electrode configurations, however, can be more sensitive to charges closer to the electrodes (see Fig.\,\ref{Fig_gas_theory}a), if the weighting potential is not linear. In an intuitive picture, this is the case, if more electric field lines of the particle charge do not end up on the electrodes but escape to the surroundings. This situation applies to all real experiments, especially solid-state experiments, where thin electrodes are deposited on the surface. As this finding suggests, the linearity of these measurements could be affected. It also implies that the scaling of the signal with electrode distance, as shown in Refs.\,\cite{Boolakee2020grapheneDistance, keiber2016novel}, will depend on the actual electrode geometry.

Additional effects might play a role in the macroscopic signal formation, such as the surface roughness of the electrodes or potentially dielectric passivation layers which could affect the weighting potentials. In future, carefully designed experiments with well characterized electrodes and modeling of the whole electric circuit, information on the weighting potentials could directly be obtained by scanning the laser focus between the electrodes and measuring both the current signal from the individual electrodes after amplification as well as the lockin-demodulated CEP-dependence. Such a characterization would have general importance to assess signal formation in ultrafast current measurements.

Despite its assumptions, our model provides surprisingly good agreement with the measurements. The model has, however, some limitations. The most important is the mean-field treatment of electrostatic interactions, that neglects electron-electron and electron-ion scattering as well as electron-ion recombination. The former two effects likely have a similar influence as electron-neutral scattering. For our experimental parameters with ionization degrees below roughly 1\,\%, this simplification seems justified. At higher intensities, however, a microscopic extension might be necessary, where additional reshaping of the laser pulse due to the generated plasma\cite{Schotz_2020_PRXphasematching} also has to be considered. Moreover, the effect of ion movement has been neglected as well as the role of retardation effects, which are important for THz generation in plasmas\cite{Cook2000THzfield, Roskos2004THzemission, Roskos2006THzCEP}. Finally, at some point plasma produced light could lead to electron emission from the electrodes, which we neglected. Although not relevant for the current study, our approach can be extended towards higher intensities by using particle-particle-particle-mesh (P$^3$M) PIC-codes, which should be able to describe most of these effects.

\section*{Conclusions}
The emergence of macroscopic currents in photoconductive sampling of optical fields has been investigated. Experimental data of the pressure dependence of the current signal amplitudes for nitrogen, argon, and air show a maximum between 10\,mbar and 100\,mbar. The electrode-distance dependence for different pressures has been evaluated and a nonlinear increase of the signal with decreasing electrode separation is found. Finally, the intensity-dependence was investigated for air and nitrogen. After a fast increase, the signal amplitude starts to saturate slightly below $1.0\cdot 10^{14}\mathrm{W/cm}^2$. The signal in air is roughly a factor 3-5 higher and starts to saturate earlier compared to nitrogen. We introduced a theoretical model for the signal generation mechanism on the electrodes based on the Ramo-Shockley-theorem. The simulations show good agreement with the experimental data indicating its validity and enabling deeper understanding of macroscopic effects in the signal formation. The model resolves the artificial separation of the dipole and current contribution that has been prevalent in the literature\cite{Kubullek2020CEPtag, Korobenko2020fsStreaking, Sederberg2020NatComm}. By comparison of the experimental results with extensive numerical particle-in-cell (PIC)-type simulations based on this model, we were able to identify the role of the electron-atom / electron-molecule scattering cross sections. We found a surprisingly large influence of the Coulomb interaction that leads to a maximum in the current signals at around 1-10\,mbar pressure for most of our experimental conditions. Our results imply that most of the heuristic models utilized so far will need to be amended. Our theoretical framework on macroscopic signal formation can be straightforwardly extended to other experimental scenarios and other media, including photo-conductive sampling of electric fields in gases and solids. Finally, as a technical advance, our findings present a way to boost the current signal measurements in gases, that have so far been done at atmospheric pressures, by more than an order of magnitude by going to lower pressures, thereby reducing scattering and charge interaction effects.

\section*{Methods}
\subsection*{Experimental details}
We used a commercial Ti:Sa laser system (Femtopower HR/CEP-4) which provides up to 0.7\,mJ pulse energy at 780\,nm with 27\,fs pulse duration at 10\,kHz repetition rate. For spectral broadening the pulses are sent through a hollow-core fiber filled with argon at 0.5\,bar. They are subsequently compressed using chirped mirrors to durations reaching 4.5\,fs in full-width-at-half-maximum (FHWM) of the intensity envelope at 750\,nm central wavelength. The temporal intensity envelope was determined via the Dispersion Scan (d-scan) technique. Pulse energies up to only 18\,$\mu$J have been used in the experiments.

For the calibration of the intensity, we measured the focal spot size inside the experimental chamber via a CCD camera. Moreover, the relative peak intensity compared to the Fourier limit for our 4.5\,fs laser pulses was determined from a d-scan measurement in front of the chamber. We obtained a conversion factor from the pulse energy, measured by a powermeter, to the peak intensity in the experimental focus of $1.1\cdot10^{14}\frac{\mathrm{W}}{\mu J\cdot\mathrm{cm}^2}\pm20\%$. For the situation including the telescope, a factor of $0.11\cdot10^{14}\frac{\mathrm{W}}{\mu J\cdot\mathrm{cm}^2}\pm50\%$ is determined. Here, a higher relative uncertainty is obtained, since the telescope introduces a slight astigmatism, affecting the accuracy of the focal spot size determination.

In order to measure the CEP-dependence of the currents, the CEP is flipped between $\varphi_\mathrm{0}$ and $\varphi_\mathrm{0}+\pi$ for consecutive laser pulses (indicated by the thick and thin red lines in Fig.\,\ref{Fig_gas_experimental_setup}a)) using an acousto-optic dispersive programmable filter (Fastlite Dazzler). Consequently, the demodulation of the voltage signals in the lock-in amplifier is performed at half the repetition rate $f_\mathrm{rep}/2$.

\subsection*{Theoretical approach}
The signal on the conducting electrodes is formed by a simple electrostatic mechanism: The induced charge $Q$ on an electrode is given by the surface charge that is induced by the presence of a charged particle $q$. The change of the induced charge $Q$ can be measured as a current if the electrodes are connected to ground. In the simplest case of an infinite conducting plate, $Q$ is given by the value of the image charge.

Immediately after ionization, the electron and parent ion are still at the same position. Since their charges have opposite signs, the induced surface charges cancel. A net charge is induced only if one charge gets displaced with respect to the other. For practical applications, it would be cumbersome to calculate the induced surface charge for each position of the electron/ion and then integrate over the electrode surface. This approach would be feasible only for very simple, highly symmetric geometries. Fortunately, the calculation is considerably simplified through the Ramo-Shockley theorem \cite{Ramo1939RSTheorem, Shockley1938RSTheorem}.

Due to computational limitations, the simulations had to be restricted to two spatial dimensions perpendicular to the laser beam propagation direction. We find this to be a good approximation because the focal spot size $\omega_\mathrm{0}$ is much smaller than the Rayleigh length $z_\mathrm{R}$. The electrons are modeled as an ensemble of pseudo-particles with an effective charge given by the total emitted charge divided by the number of pseudo-particles $N$. For the results presented here, we use $N=5\cdot10^5$. The total charge is obtained by radially integrating the final ionization fraction calculated by the ADK rate\cite{Tonglin2005JourPhysB} multiplied by the atomic number density ($\propto$ pressure $p$). For nitrogen ($I_\mathrm{p}$=\,15.58\,eV) the same tunneling rate as for argon ($I_\mathrm{p}$=\,15.76\,eV) is employed. In order to model the contribution of oxygen in air, which has a much lower ionization potential ($I_\mathrm{p}$=\,12.56\,eV)\cite{Lin2002O2vsXe} than nitrogen, we use the ADK-parameters of xenon ($I_\mathrm{p}$=\,12.13\,eV), but with angular momentum quantum numbers of $l$=2 and $m$=1. The latter is important since it takes into account the symmetry of the molecular wavefunction of O$_\mathrm{2}$ in the tunneling region which leads to a significantly lower tunneling rate than in xenon ($l$=1, $m$=0)\cite{Lin2002O2vsXe}.

In our Monte-Carlo approach, the initial position of a pseudo-electron and the corresponding pseudo-ion is randomly sampled from the radially-resolved ionization fraction and the emission time from the tunneling rate. From the latter, the final emission velocity of the pseudo-electron is calculated in the SMM within the SFA under the assumption that effectively only direct electrons contribute. Pseudo-electrons have a charge-to-mass ratio of $e/m_\mathrm{e}$ such that they behave like normal electrons during propagation. Pseudo-ions are assumed to stay fixed at the birth position.

The propagation of pseudo-electrons is performed via the Velocity-Verlet algorithm\cite{Swope1982VelocityVerlet} using a time-step of 20\,fs, or smaller if required, over a time-span of 1\,ns. For each time step, the electron-neutral (atom or molecule) scattering probability is calculated via the mean-free path $l_\mathrm{mfp}$ and Monte-Carlo sampling. $l_\mathrm{mfp}$ is obtained from the MagBoltz\cite{Magboltz} cross-sections available via the xcat-database\cite{Biagi_database}, that contain elastic, excitation and ionization cross-sections. The mean-free paths for argon (red line) and nitrogen (blue line) at 1\,mbar are shown in Fig.\,\ref{Fig_gas_theory}b). The mean-free-path in both argon and nitrogen are similar above 10 eV; however, it increases in argon above the threshold of ionization and excitation. Differences between the two gases reach more than an order of magnitude at electron kinetic energies around 0.5\,eV. In nitrogen, on the other hand, excitation channels at lower energies due to the molecular structure result in a minimum in $l_\mathrm{mfp}$ at around 2\,eV. It is important to note that the mean-free path scales with $1/p$, so $l_\mathrm{mfp}$ is on the order of 1\,mm for 1\,mbar, while for 1000\,mbar it is around 1\,$\mu$m. The scattering time is above $10^{-13}$\,s even at atmospheric pressure. For the inelastic channels we assume a uniform probability for the energy loss from the threshold of the inelastic process, e.g. the ionization potential for the ionization channel, up to the electron kinetic energy. Secondary electrons are neglected. For simplicity, we assume isotropic scattering in the lab frame for all processes. Since we deal with ionization fractions of around 1\,\% and below, scattering and recombination with the ions is neglected. For air, the scattering cross-sections of nitrogen are used.

In order to calculate the electrostatic interaction, the Poisson equation is solved on a 2D-grid for each time step. The length of the rectangular simulation region in the $x$-direction is the electrode distance $D$. Along the $y$-direction, the length is taken to be $3 D$. The grid contains 512 points along the electrode axis and 1536 along the other axis. The laser focus is positioned in the center of the rectangle. The pseudo-electrons are sampled on the grid using a linear weighting scheme. We impose the Dirichlet boundary condition for the potential $\phi=0$ at all four edges of the simulation region using the method of image charges. Towards this end, the grid is doubled in size and a charge of opposite sign is injected at the position mirrored along the positive edge. Due to the implicit periodicity of the Fast-Fourier Transform (FFT) used for solving the Poisson equation, all contributions of the otherwise infinite sum of mirror charges are contained in the calculation. The electric field is obtained in the same step and linearly interpolated onto the positions of the pseudo-electrons. If a particle leaves the simulation region it is not considered anymore in the electrostatic interaction. Across the boundary perpendicular to the electrodes the propagation is continued whereas it is stopped if the pseudo-electron reaches the electrodes. For the pseudo-ions, the potential and field calculation on the grid is only calculated once at the start of the simulation.

At each time step, the induced charge $Q$ on both electrodes is calculated using the linear weighting potential of the infinite parallel plates shown above and summing over the ensemble of pseudo-electrons. Since the induced current decays over a timescale of 100\,fs -- 1\,ps (see discussion below), much faster than the bandwidth of measurement electronics, the measured signal is assumed to be proportional to the induced charge $Q$ averaged from 0.5\,ns -- 1\,ns after the start of the simulation, when $Q$ has reached a quasi asymptotic value (see Fig.\,\ref{Fig_gas_scatteringCoulomb}a)). The lower bound does, however, not impact the results significantly. In order to obtain the signal in the experiment from the 2D-simulation, the simulated induced charge density is multiplied by the repetition rate of the laser (10\,kHz), the transimpedance gain ($10^9$\,V/A) and the effective ionization length $\Delta z_\mathrm{ion,eff}$ which is a free parameter, whose order of magnitude should be between the focal spotsize and the Rayleigh length. The used values are $\Delta z_\mathrm{ion,eff}=7\mu$m (Fig.\,\ref{Fig_gas_pressure_results}\,b)), $\Delta z_\mathrm{ion,eff}=44\,\mu$m (Fig.\,\ref{Fig_gas_distance_experiment}), $\Delta z_\mathrm{ion,eff}=47\,\mu\mathrm{m}$ (Fig.\,\ref{Fig_gas_intensity_experiment}) and $\Delta z_\mathrm{ion,eff}=25\,\mu\mathrm{m}$ (Fig.\,\ref{Fig_gas_scatteringCoulomb}\,c)).

\section*{Acknowledgements}
We acknowledge support by the German Research Foundation (DFG) via SFB NOA and LMUexcellent, by the European Union via FETopen PetaCOM and FETlaunchpad FIELDTECH, and by the Max Planck Society via the IMPRS for Advanced Photon Science and the Max-Planck Fellow program. We acknowledge support by the King-Saud university in the framework of the MPQ-KSU collaboration. We thank Nicholas Karpowicz for fruitful discussions, and are grateful for support by Ferenc Krausz.

\section*{Author Contributions}
M.W., V.Y., B.B., and M.F.K. directed the project. J.S., A.M., J.B. and D.Z. carried out the experiments and analyzed the data. J.S. developed the theoretical model and performed all simulations. P.R. helped with operating the laser. Z.W. contributed to the implementation of the experimental setup and the discussion. M.A. and A.M.A. contributed to the discussion. J.S. wrote the manuscript, which was finalized with input from all authors.

\section*{Competing interests}
The authors declare no competing interests.


\end{document}